\documentclass[a4paper,11pt]{article}
\pdfoutput=1

\usepackage{jheppub}
\usepackage{amsmath}

\newcommand{\bea}{\begin{eqnarray}}
\newcommand{\eea}{\end{eqnarray}}

\makeatletter
\@addtoreset{equation}{section}
\makeatother

\title{
Chiral Ground States in a Frustrated Holographic Superconductor
}
\author{
Mitsuhiro Nishida
}

\affiliation
{Department of Physics, Graduate School of Science,
Osaka University, Toyonaka, Osaka 560-0043, Japan}

\emailAdd{nishida(at)het.phys.sci.osaka-u.ac.jp}

\abstract{Frustration is an important phenomenon in condensed matter physics because it can introduce a new order parameter such as chirality. Towards understanding a mechanism of the frustration in strongly correlated systems, we study a holographic superconductor model with three scalar fields and an interband Josephson coupling, which is important for the frustration. We analyze free energy of solutions of the model to determine ground states. We find chiral ground states, which have nonzero chirality.
}

\preprint{OU-HET 841
}

\begin{document}
\maketitle
\setcounter{page}{1}

\section{Introduction}
\label{sec1}
Frustration is widely studied in condensed matter physics (see, for example, \cite{Moessner,Diep}). This frustration means a situation where several constraints compete and it can cause degeneracy of vacuum.  One famous example is antiferromagnets on a triangular lattice. Some frustrated systems have a characteristic order parameter such as chirality, which represents a property under a mirror operation. There is a strange phenomenon by nonzero chirality, for example, anomalous Hall effect \cite{Ohgushi}. Study of the frustration is aimed to create new materials by using a property of frustrated systems.

In condensed matter physics, there is a three-band superconductor model in which the frustration occurs \cite{Tanaka}. This model has three interband Josephson coupling terms between three scalar fields and they lead the frustration. Furthermore, chiral ground states\footnote{In this paper, ground state means the lowest energy state at finite temperature.}exist in this model. These chiral ground states are vacua, which have nonzero chirality and their origin is the frustration. 

In this paper, we find a holographic superconductor model \cite{HHH} in which the frustration occurs. From a viewpoint of the gauge/gravity correspondence \cite{Maldacena:1997re, Gubser:1998bc, Witten:1998qj}, holographic superconductor models with several order parameters are well-studied \cite{Basu:2010fa, Wen:2010et, Huang:2011ac, Zayas:2011dw, Krikun:2012yj, Musso:2013ija, Cai:2013wma, Nitti:2013xaa, Liu:2013yaa, Nie:2013sda, Amado:2013lia, Amoretti:2013oia, Donos:2013woa, Nishida:2014lta, Li:2014wca, Nitti:2014fsa, Chaturvedi:2014dga, Nie:2014qma}. In particular, a two-scalar model with the Josephson coupling was studied in Ref.~\cite{Wen:2013ufa}. However, we need more than two scalar fields for the frustration since we need some constraints of complex phases of the scalar fields. Therefore, we consider a three-scalar model with the Josephson coupling for the frustration. In order to determine ground states, we analyze free energy of solutions of this model. In Ref.~\cite{Tanaka}, the authors introduce the Josephson coupling in order to obtain the chiral ground states. In our holographic model, we first consider the three-scalar model with the Josephson coupling. Next, we add a quartic potential term in order to obtain the chiral ground states. When we choose specific values of parameters, there are solutions corresponding to the chiral ground states. This model may be a help to understand a mechanism of the frustration in strongly correlated systems from the viewpoint of the gauge/gravity correspondence.

The organization of this paper is as follows. In section 2, we analyze a holographic two-scalar model and see that the frustration does not occur in this model. In section 3, we analyze a holographic three-scalar model and see that there are the chiral ground states by comparing the free energy when we choose specific values of parameters. Section 4 is a summary and a discussion.

\section{Two-scalar model}
\label{sec2}
In this section, we study a two-scalar model as a minimal model with the Josephson coupling. This model was studied first in Ref.~\cite{Wen:2013ufa}\footnote{In Ref.~\cite{Wen:2013ufa}, effect of the Josephson coupling to boundary conditions of the scalar fields at the AdS boundary is not considered. We will derive this effect by diagonalization.}. We see that the two-scalar model can be solved and cannot describe the frustration.

In this paper, we use a four-dimensional AdS planar black hole space-time 
\begin{align}
ds^2=\frac{L^2}{z^2}(-f(z)&dt^2+dx^2+dy^2+\frac{dz^2}{f(z)}),\\
f(z)&=1-\Big(\frac{z}{z_h}\Big)^3,
\end{align}
where $z=0$ is the AdS boundary and $z=z_h$ is the horizon of the black hole. This metric is often used in the field of the holographic superconductor since a phase transition of the order parameter occurs by changing the Hawking temperature of the black hole. For simplicity, we fix the metric and use the probe limit.

For a two-scalar model with the Josephson coupling, we consider an action with a Maxwell field $A_\mu$, two complex scalar fields $\varphi_i$ and the nonzero Josephson coupling $\epsilon$ as
\begin{align}
S=\int d^4x\sqrt{-g}\Big[&-\frac{1}{4}F^{\mu\nu}F_{\mu\nu}-|D_\mu\varphi_1|^2-|D_\mu\varphi_2|^2\notag\\
&-m_1^2|\varphi_1|^2-m_2^2|\varphi_2|^2-\epsilon(\varphi_1^*\varphi_2+\varphi_1\varphi_2^*)\Big],\\
D_\mu=\nabla_\mu-iA_\mu.\;\;\;&
\end{align}
Because of the existence of the Josephson coupling, phase angles of the scalar fields are constrained as we will see in (2.14). 

The equations of motion of $\varphi_i$ are 
\begin{align}
D^\mu D_\mu\varphi_1-m^2_1\varphi_1-\epsilon\varphi_2=0,\\
D^\mu D_\mu\varphi_2-m^2_2\varphi_2-\epsilon\varphi_1=0.
\end{align}
In these equations, two types of the solution are possible:
\begin{itemize}
\item $\varphi_1=\varphi_2=0.$ This solution corresponds to the normal conducting phase.
\item $\varphi_1\neq0, \varphi_2\neq0.$ This solution corresponds to the superconducting phase.
\end{itemize}
If $\varphi_1=0$, we obtain $\varphi_2=0$ from (2.5). Therefore, there are no solutions as $\varphi_1\neq0, \varphi_2=0$ and $\varphi_1=0, \varphi_2\neq0$.

From now, we consider the solution with $\varphi_1\neq0, \varphi_2\neq0$ to describe the superconducting phase. This solution exists when the Hawking temperature of the black hole is low enough. In this case, we can rewrite $\varphi_i$ by absolute values $\psi_i$ and phase angles $\theta_i$ as
\begin{align}
\varphi_i=\psi_ie^{i\theta_i},
\end{align}
where $\psi_i>0$. For homogeneousness of the fields and simplicity of $\theta_i$, our ansatz for the fields is 
\begin{align}
A_t=A_t(z),\;\;\;\;\;\psi_i=\psi_i(z),\;\;\;\;\;\theta_i=\textrm{const}.,
\end{align}
and other components are zero. Under this ansatz, the equations of motion are
\begin{align}
&\nabla_\mu F^{\mu\nu}-2\psi^2_1A^\nu-2\psi_2^2A^\nu=0,\\
&\nabla_\mu\nabla^\mu\psi_1-A_\mu A^\mu\psi_1-m^2_1\psi_1-\epsilon\psi_2\cos{(\theta_1-\theta_2)}=0,\\
&\nabla_\mu\nabla^\mu\psi_2-A_\mu A^\mu\psi_2-m^2_2\psi_2-\epsilon\psi_1\cos{(\theta_1-\theta_2)}=0,\\
&\epsilon\psi_1\psi_2\sin{(\theta_1-\theta_2)}=0,\\
&\epsilon\psi_1\psi_2\sin{(\theta_2-\theta_1)}=0.
\end{align}

From the equations of motion of $\theta_i$, we obtain
\begin{align}
\theta_1-\theta_2=0\;\;\textrm{or}\;\;\pi .
\end{align}
This solution is not frustrated because it minimizes the coefficient of the Josephson coupling potential term $2\epsilon\cos{(\theta_1-\theta_2)}$ in (2.3). Therefore, the two-scalar model cannot describe the frustration. To describe the frustration, we need more than two scalar fields as section 3.

Incidentally, the two-scalar model can be solved by diagonalization.
To rewrite the equations of motion of $\psi_i$ by a diagonal basis of the scalar fields, we define $\epsilon', \lambda_i$ and $\psi_i'$ as 
\begin{align}
\epsilon'&\equiv\epsilon\cos{(\theta_1-\theta_2)},\\
\lambda_1&\equiv\frac{(m^2_1+m^2_2)-\sqrt{(m_1^2-m_2^2)^2+4\epsilon'^2}}{2},\\
\lambda_2&\equiv\frac{(m^2_1+m^2_2)+\sqrt{(m_1^2-m_2^2)^2+4\epsilon'^2}}{2},\\
\psi'_1&\equiv\frac{-\epsilon'\psi_1+(m^2_1-\lambda_1)\psi_2}{\sqrt{\epsilon'^2+(m_1^2-\lambda_1)^2}},\\
\psi'_2&\equiv\frac{(m^2_1-\lambda_1)\psi_1+\epsilon'\psi_2}{\sqrt{\epsilon'^2+(m_1^2-\lambda_1)^2}}.
\end{align}
Using them, the equations of motion of $A_\mu$ and $\psi_i'$ are written as
\begin{align}
&\nabla_\mu F^{\mu\nu}-2\psi'^2_1A^\nu-2\psi'^2_2A^\nu=0,\\
&\nabla_\mu\nabla^\mu\psi_1'-A_\mu A^\mu\psi_1'-\lambda_1\psi_1'=0,\\
&\nabla_\mu\nabla^\mu\psi_2'-A_\mu A^\mu\psi_2'-\lambda_2\psi_2'=0.
\end{align}
These equations of motion are same as those of a two-scalar model without the Josephson coupling. We assume that there is no solution with $\psi'_1\neq0, \psi'_2\neq0$ since it was checked that there is not such solution if charges of the scalar fields are same and $\lambda_1\neq\lambda_2$ by numerical calculations in Ref.~\cite{Basu:2010fa}.  

For example, we consider the solution in the case of $\epsilon'<0$. In this case, the solution is $\psi'_1\neq0, \psi'_2=0$ since $-\epsilon'\psi_1+(m^2_1-\lambda_1)\psi_2$ is positive and we obtain
\begin{align}
&\nabla_\mu F^{\mu\nu}-2\psi'^2_1A^\nu=0,\\
&\nabla_\mu\nabla^\mu\psi_1'-A_\mu A^\mu\psi_1'-\lambda_1\psi_1'=0,\\
&\psi'_1=-\frac{\sqrt{\epsilon'^2+(m_1^2-\lambda_1)^2}}{\epsilon'}\psi_1=\frac{\sqrt{\epsilon'^2+(m_1^2-\lambda_1)^2}}{m_1^2-\lambda_1}\psi_2.
\end{align}
(2.23) and (2.24) are same as the equations of motion of a one-scalar model \cite{HHH}. It is well-known that the solution with $\psi'_1\neq0$ exists if the Hawking temperature of the black hole is low enough. Therefore, we conclude that the two-scalar model can be solved. Similarly, we can obtain the solution in the case of $\epsilon'>0$.

In the gauge/gravity correspondence, the free energy is related to on-shell Euclidean action. Generally, the free energy of the solution in the case of $\epsilon'<0$ is smaller than that of $\epsilon'>0$ since $\lambda_1<\lambda_2$. It is reasonable because $\epsilon'$ is a coefficient of the Josephson coupling potential term in (2.3). From (2.24), the power of $z$ in $\psi'_1$ at the AdS boundary is determined as
\begin{align}
&\psi'_1=\psi'^{(1)}_1z^{\Delta_1}+\psi'^{(2)}_2z^{\Delta_2},\\
&\Delta_1=\frac{3-\sqrt{9+4\lambda_1}}{2},\;\;\Delta_2=\frac{3+\sqrt{9+4\lambda_1}}{2}.
\end{align}
If $|\epsilon|$ is large enough, $\lambda_1$ is below Breitenlohner-Freedman bound \cite{Breitenlohner:1982jf} and the theory is unstable.

\section{Three-scalar model}
\label{sec3}
In this section, we study a three-scalar model as a model in which the frustration occurs. When we choose specific values of parameters, the frustration is realized and the solutions which correspond to the chiral ground states exist.  

For a three-scalar model with the Josephson coupling, we consider an action with a Maxwell field $A_\mu$, three complex scalar fields $\varphi_i$, three nonzero Josephson coupling $\epsilon_{ij}$ and a nonnegative constant $\eta$ as
\begin{align}
S=\int d^4x\sqrt{-g}\Big[&-\frac{1}{4}F^{\mu\nu}F_{\mu\nu}-|D_\mu\varphi_1|^2-|D_\mu\varphi_2|^2-|D_\mu\varphi_3|^2\notag\\
&-m_1^2|\varphi_1|^2-m_2^2|\varphi_2|^2-m_3^2|\varphi_3|^2\notag\\
&-\epsilon_{12}(\varphi_1^*\varphi_2+\varphi_1\varphi_2^*)-\epsilon_{23}(\varphi_2^*\varphi_3+\varphi_2\varphi_3^*)-\epsilon_{31}(\varphi_3^*\varphi_1+\varphi_3\varphi_1^*)\notag\\
&-\eta(|\varphi_1|^4+|\varphi_2|^4+|\varphi_3|^4)\Big].
\end{align}
If $\epsilon_{12}\epsilon_{12}\epsilon_{31}>0$, there is no configuration of $\theta_i$ which minimizes each coefficient of the Josephson coupling potential terms. In this case, the chiral ground states can exist. The quartic potential of this action is symmetric about $\varphi_i$ and it is one example of the potential that the chiral ground states can exist. We use the metric (2.1) and the probe limit for simple calculation. 

\subsection{Analysis of the solutions with $\eta=0$}

First, we ignore the quartic potential and analyze the equations of motion with $\eta=0$. The equations of motion of $\varphi_i$ are
\begin{align}
D^\mu D_\mu\varphi_1-m^2_1\varphi_1-\epsilon_{12}\varphi_2-\epsilon_{31}\varphi_3=0,\\
D^\mu D_\mu\varphi_2-m^2_2\varphi_2-\epsilon_{23}\varphi_3-\epsilon_{12}\varphi_1=0,\\
D^\mu D_\mu\varphi_3-m^2_3\varphi_3-\epsilon_{31}\varphi_1-\epsilon_{23}\varphi_2=0.
\end{align}
In these equations, three types of the solution are possible:
\begin{enumerate}
\item[Sol.1] $\varphi_1=\varphi_2=\varphi_3=0.$ 
\item[Sol.2] One scalar field is zero and others are nonzero as $\varphi_1\neq0, \varphi_2\neq0, \varphi_3=0$.
\item[Sol.3] $\varphi_1\neq0, \varphi_2\neq0, \varphi_3\neq0$.
\end{enumerate}

To analyze the frustration between three scalar fields, we consider the solution 3. We rewrite $\varphi_i$ as (2.7) and use the ansatz as (2.8) in the same way of section 2. Under this ansatz, the equations of motion are
\begin{align}
&\nabla_\mu F^{\mu\nu}-2\psi^2_1A^\nu-2\psi_2^2A^\nu-2\psi^2_3A^\nu=0,\\
&\nabla_\mu\nabla^\mu\psi_1-A_\mu A^\mu\psi_1-m^2_1\psi_1-\epsilon'_{12}\psi_2-\epsilon'_{31}\psi_3=0,\\
&\nabla_\mu\nabla^\mu\psi_2-A_\mu A^\mu\psi_2-m^2_2\psi_2-\epsilon'_{23}\psi_3-\epsilon'_{12}\psi_1=0,\\
&\nabla_\mu\nabla^\mu\psi_3-A_\mu A^\mu\psi_3-m^2_3\psi_3-\epsilon'_{31}\psi_1-\epsilon'_{23}\psi_2=0,\\
&\epsilon_{12}\psi_1\psi_2\sin{(\theta_1-\theta_2)}+\epsilon_{31}\psi_1\psi_3\sin{(\theta_1-\theta_3)}=0,\\
&\epsilon_{23}\psi_2\psi_3\sin{(\theta_2-\theta_3)}+\epsilon_{12}\psi_2\psi_1\sin{(\theta_2-\theta_1)}=0,\\
&\epsilon_{31}\psi_3\psi_1\sin{(\theta_3-\theta_1)}+\epsilon_{23}\psi_3\psi_2\sin{(\theta_3-\theta_2)}=0,\\
&\epsilon'_{12}\equiv\epsilon_{12}\cos{(\theta_1-\theta_2)}, \;\;\;\epsilon'_{23}\equiv\epsilon_{23}\cos{(\theta_2-\theta_3)}, \;\;\;\epsilon'_{31}\equiv\epsilon_{31}\cos{(\theta_3-\theta_1)}.
\end{align}
In the equations of $\theta_i$, two types of the solution are possible:
\begin{enumerate}
\item[Sol.3a] $\sin{(\theta_1-\theta_2)}\neq0, \sin{(\theta_2-\theta_3)}\neq0, \sin{(\theta_3-\theta_1)}\neq0.$ \item[Sol.3b] $\sin{(\theta_1-\theta_2)}=\sin{(\theta_2-\theta_3)}=\sin{(\theta_3-\theta_1)}=0.$  
\end{enumerate}
In the three-scalar model, there is the solution 3a with $\sin{(\theta_i-\theta_j)}\neq0$ unlike the two-scalar model. This solution has nonzero chirality.

We derive a condition that the solution 3a is possible. From (3.10) and (3.11), we obtain
\begin{align}
\psi_2=-\frac{\epsilon_{31}\sin{(\theta_3-\theta_1)}}{\epsilon_{23}\sin{(\theta_3-\theta_2)}}\psi_1,\;\;
\psi_3=-\frac{\epsilon_{12}\sin{(\theta_2-\theta_1)}}{\epsilon_{23}\sin{(\theta_2-\theta_3)}}\psi_1.
\end{align}
Since $\psi_i>0$, we need
\begin{align}
\frac{\epsilon_{31}\sin{(\theta_3-\theta_1)}}{\epsilon_{23}\sin{(\theta_3-\theta_2)}}<0,\;\;\;\frac{\epsilon_{12}\sin{(\theta_2-\theta_1)}}{\epsilon_{23}\sin{(\theta_2-\theta_3)}}<0.
\end{align}
Substituting (3.13) for (3.6), (3.7) and (3.8), we obtain
\begin{align}
&\nabla_\mu\nabla^\mu\psi_1-A_\mu A^\mu\psi_1-\left(m^2_1-\frac{\epsilon_{12}\epsilon_{31}}{\epsilon_{23}}\right)\psi_1=0,\\
&\nabla_\mu\nabla^\mu\psi_1-A_\mu A^\mu\psi_1-\left(m^2_2-\frac{\epsilon_{23}\epsilon_{12}}{\epsilon_{31}}\right)\psi_1=0,\\
&\nabla_\mu\nabla^\mu\psi_1-A_\mu A^\mu\psi_1-\left(m^2_3-\frac{\epsilon_{31}\epsilon_{23}}{\epsilon_{12}}\right)\psi_1=0.
\end{align}
Therefore, the solution 3a is possible only if
\begin{align}
m^2_1-\frac{\epsilon_{12}\epsilon_{31}}{\epsilon_{23}}=m^2_2-\frac{\epsilon_{23}\epsilon_{12}}{\epsilon_{31}}=m^2_3-\frac{\epsilon_{31}\epsilon_{23}}{\epsilon_{12}}.
\end{align}

To compare the free energy of the solutions, we analyze other solutions with (3.18). 

\begin{enumerate}
\item[Sol.1] \underline{$\varphi_1=\varphi_2=\varphi_3=0.$} 

Generally, the free energy of the solution with $\varphi_1=\varphi_2=\varphi_3=0$ is larger than that with the nonzero scalar fields. Therefore, this solution is not a ground state if the Hawking temperature of the black hole is low enough.

\item[Sol.2] \underline{$\varphi_1\neq0, \varphi_2\neq0, \varphi_3=0$.}

From (3.4), we obtain
\begin{align}
\psi_2=-\frac{\epsilon_{31}}{\epsilon_{23}}\psi_1e^{i(\theta_1-\theta_2)},
\end{align}
and 
\begin{align}
\theta_1-\theta_2=0\;\;\textrm{or}\;\;\pi,
\end{align}
are the solutions. Substituting (3.19) for (3.6) and (3.7), we obtain
\begin{align}
&\nabla_\mu\nabla^\mu\psi_1-A_\mu A^\mu\psi_1-\left(m^2_1-\frac{\epsilon_{12}\epsilon_{31}}{\epsilon_{23}}\right)\psi_1=0,\\
&\nabla_\mu\nabla^\mu\psi_1-A_\mu A^\mu\psi_1-\left(m^2_2-\frac{\epsilon_{23}\epsilon_{12}}{\epsilon_{31}}\right)\psi_1=0,
\end{align}
and the solution with $\varphi_1\neq0, \varphi_2\neq0, \varphi_3=0$ is possible if (3.18) holds. Similarly, other solutions with $\varphi_1=0, \varphi_2\neq0, \varphi_3\neq0$ and $\varphi_1\neq0, \varphi_2=0, \varphi_3\neq0$ are possible. Free energy of these solutions is same as that of solution 3a because the mass squared of $\psi_i$ in (3.15) and (3.21) is same.

\item[Sol.3b] \underline{$\sin{(\theta_1-\theta_2)}=\sin{(\theta_2-\theta_3)}=\sin{(\theta_3-\theta_1)}=0.$}

In this case, the diagonalization as section 2 is useful. We diagonalize a matrix as
\begin{align}
\begin{pmatrix}
m^2_1&\epsilon'_{12}&\epsilon'_{31}\\
\epsilon'_{12}&m^2_2&\epsilon'_{23}\\
\epsilon'_{31}&\epsilon'_{23}&m^2_3
\end{pmatrix}.
\end{align}
If (3.18) holds, this matrix can be transformed to a diagonal matrix as
\begin{align}
\begin{pmatrix}
m^2_1-\frac{\epsilon_{12}\epsilon_{31}}{\epsilon_{23}}&0&0\\
0&m^2_1-\frac{\epsilon_{12}\epsilon_{31}}{\epsilon_{23}}&0\\
0&0&m^2_1+\frac{\epsilon_{31}\epsilon_{23}}{\epsilon_{12}}+\frac{\epsilon_{23}\epsilon_{12}}{\epsilon_{31}}
\end{pmatrix}.
\end{align}
If $\epsilon_{12}\epsilon_{23}\epsilon_{31}>0$, a minimum value of the free energy of this solution is same as that of the solution 3a since $m^2_1-\frac{\epsilon_{12}\epsilon_{31}}{\epsilon_{23}}<m^2_1+\frac{\epsilon_{31}\epsilon_{23}}{\epsilon_{12}}+\frac{\epsilon_{23}\epsilon_{12}}{\epsilon_{31}}$\footnote{We assume that there are the solutions as those of a one-scalar model only as section 2. We will explain this assumption later.}. However, if  $\epsilon_{12}\epsilon_{23}\epsilon_{31}<0$, the free energy of the solution corresponds to $m^2_1+\frac{\epsilon_{31}\epsilon_{23}}{\epsilon_{12}}+\frac{\epsilon_{23}\epsilon_{12}}{\epsilon_{31}}$ is smaller than that of the solution 3a and the frustration does not occur.
\end{enumerate}

Summarizing the above, if $\eta=0$, $\epsilon_{12}\epsilon_{23}\epsilon_{31}>0$ and (3.18) hold, there are several solutions whose free energy is same by the frustration.

\subsection{Analysis of the free energy with $\eta>0$}

Next, we consider the three-scalar model with $\eta>0$, $\epsilon_{12}\epsilon_{23}\epsilon_{31}>0$ and (3.18). Substituting the solutions with $\eta>0$ for the action, we analyze the free energy with $\eta>0$. The equations of motion of $\psi_i$ are 
\begin{align}
&\nabla_\mu\nabla^\mu\psi_1-A_\mu A^\mu\psi_1-m^2_1\psi_1-\epsilon'_{12}\psi_2-\epsilon'_{31}\psi_3-2\eta\psi_1^3=0,\\
&\nabla_\mu\nabla^\mu\psi_2-A_\mu A^\mu\psi_2-m^2_2\psi_2-\epsilon'_{23}\psi_3-\epsilon'_{12}\psi_1-2\eta\psi_2^3=0,\\
&\nabla_\mu\nabla^\mu\psi_3-A_\mu A^\mu\psi_3-m^2_3\psi_3-\epsilon'_{31}\psi_1-\epsilon'_{23}\psi_2-2\eta\psi_3^3=0,
\end{align}
and the equations of motion of $A_\mu$ and $\theta_i$ are (3.5), (3.9), (3.10) and (3,11).
For simplicity, we set 
\begin{align}
 m^2_1=m^2_2=m^2_3, \;\;\epsilon_{12}=\epsilon_{23}=\epsilon_{31}>0.
\end{align}
In this case, difference of $\theta_i$ of the chiral ground states becomes symmetric, but we lose generality of the model.

\begin{enumerate}
\item[Sol.2] \underline{$\varphi_1\neq0, \varphi_2\neq0, \varphi_3=0.$}

Substituting (3.19) to (3.25) and (3.26), we obtain
\begin{align}
&\nabla_\mu\nabla^\mu\psi_1-A_\mu A^\mu\psi_1-\left(m^2_1-\frac{\epsilon_{12}\epsilon_{31}}{\epsilon_{23}}\right)\psi_1-2\eta\psi_1^3=0,\\
&\nabla_\mu\nabla^\mu\psi_1-A_\mu A^\mu\psi_1-\left(m^2_2-\frac{\epsilon_{23}\epsilon_{12}}{\epsilon_{31}}\right)\psi_1-2\eta\left(\frac{\epsilon_{31}}{\epsilon_{23}}\right)^2\psi_1^3=0,
\end{align}
and these equations are same if (3.28) holds. This equation is the equation of one-scalar model with a quartic potential in the probe limit \cite{Gubser:2008wz, Gubser:2009cg}. Substituting (3.19) to (3.1), we obtain the on-shell action

\begin{align}
S_{\textrm{on-shell}}=\int d^4x\sqrt{-g}\Big[&-\frac{1}{4}F^{\mu\nu}F_{\mu\nu}-2|D_\mu\psi_1|^2-2\left(m_1^2-\frac{\epsilon_{12}\epsilon_{31}}{\epsilon_{23}}\right)\psi_1^2
-2\eta\psi_1^4\Big]\notag\\
=\int d^4x\sqrt{-g}\Big[&-\frac{1}{4}F^{\mu\nu}F_{\mu\nu}-|D_\mu\psi'|^2-\left(m_1^2-\frac{\epsilon_{12}\epsilon_{31}}{\epsilon_{23}}\right)\psi'^2-\frac{\eta}{2}\psi'^4\Big],
\end{align}
where we redefine a new scalar field $\psi'$ to compare the free energy as
\begin{align}
\psi'^2\equiv2\psi_1^2.
\end{align}
In this solution, a coefficient of the quartic potential of $\psi'$ is $\eta/2$.
The solutions with $\varphi_1=0, \varphi_2\neq0, \varphi_3\neq0$ and $\varphi_1\neq0, \varphi_2=0, \varphi_3\neq0$ are also the same.

\item[Sol.3a] \underline{$\sin{(\theta_1-\theta_2)}\neq0, \sin{(\theta_2-\theta_3)}\neq0, \sin{(\theta_3-\theta_1)}\neq0.$} 

Substituting (3.13) to (3.25), (3.26) and (3.27), we obtain
\begin{align}
&\nabla_\mu\nabla^\mu\psi_1-A_\mu A^\mu\psi_1-\left(m^2_1-\frac{\epsilon_{12}\epsilon_{31}}{\epsilon_{23}}\right)\psi_1-2\eta\psi_1^3=0,\\
&\nabla_\mu\nabla^\mu\psi_1-A_\mu A^\mu\psi_1-\left(m^2_2-\frac{\epsilon_{23}\epsilon_{12}}{\epsilon_{31}}\right)\psi_1-2\eta\left(\frac{\epsilon_{31}\sin{(\theta_3-\theta_1)}}{\epsilon_{23}\sin{(\theta_3-\theta_2)}}\right)^2\psi_1^3=0,\\
&\nabla_\mu\nabla^\mu\psi_1-A_\mu A^\mu\psi_1-\left(m^2_3-\frac{\epsilon_{31}\epsilon_{23}}{\epsilon_{12}}\right)\psi_1-2\eta\left(\frac{\epsilon_{12}\sin{(\theta_2-\theta_1)}}{\epsilon_{23}\sin{(\theta_2-\theta_3)}}\right)^2\psi_1^3=0.
\end{align}
If (3.28) is satisfied, the solution 3a exists only if
\begin{align}
1=\left(\frac{\sin{(\theta_3-\theta_1)}}{\sin{(\theta_3-\theta_2)}}\right)^2=
\left(\frac{\sin{(\theta_2-\theta_1)}}{\sin{(\theta_2-\theta_3)}}\right)^2.
\end{align}

In the range of (3.14), there are two configurations of $\theta_i$ which satisfy (3.36):
\begin{align}
\theta_1-\theta_2=\theta_2-\theta_3=\theta_3-\theta_1=\frac{2\pi}{3},
\end{align}
and
\begin{align}
\theta_1-\theta_2=\theta_2-\theta_3=\theta_3-\theta_1=\frac{4\pi}{3}.
\end{align}
These solutions are chiral as figure 1. We can define a physical quantity to determine the chirality. For example, the sign of $i(\varphi_1^*\varphi_2-\varphi_1\varphi_2^*)=2\psi_1\psi_2\sin{(\theta_1-\theta_2)}$ is different for each solution. Substituting (3.13) and (3.36) to (3.1), we obtain the on-shell action

\begin{align}
S_{\textrm{on-shell}}=\int d^4x\sqrt{-g}\Big[&-\frac{1}{4}F^{\mu\nu}F_{\mu\nu}-3|D_\mu\psi_1|^2-3\left(m_1^2-\frac{\epsilon_{12}\epsilon_{31}}{\epsilon_{23}}\right)\psi_1^2-3\eta\psi_1^4\Big]\notag\\
=\int d^4x\sqrt{-g}\Big[&-\frac{1}{4}F^{\mu\nu}F_{\mu\nu}-|D_\mu\psi'|^2-\left(m_1^2-\frac{\epsilon_{12}\epsilon_{31}}{\epsilon_{23}}\right)\psi'^2-\frac{\eta}{3}\psi'^4\Big],
\end{align}
where we redefine a new scalar field $\psi'$ as
\begin{align}
\psi'^2\equiv3\psi_1^2,
\end{align}
and the coefficient of the quartic potential of $\psi'$ is $\eta/3$.

\begin{figure}[t]
\centering
\includegraphics[scale=0.4]{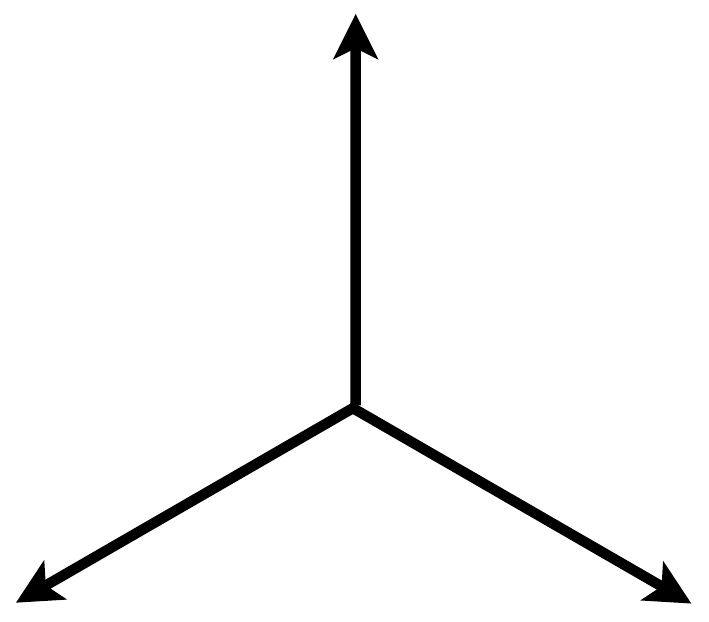}\hspace{10mm}
\includegraphics[scale=0.4]{figure2.pdf}
\put(-158,75){$\theta_1$}
\put(-44,75){$\theta_1$}
\put(-112,-6){$\theta_2$}
\put(-88,-6){$\theta_2$}
\put(-204,-6){$\theta_3$}
\put(-0,-6){$\theta_3$}
\caption{Chiral solutions (3.37) and (3.38). Three allows represent the phase angles of the scalar fields. These solutions interchange by a mirror operation.}
\label{fig1}
\end{figure}

\item[Sol.3b] \underline{$\sin{(\theta_1-\theta_2)}=\sin{(\theta_2-\theta_3)}=\sin{(\theta_3-\theta_1)}=0.$}

For example, we consider the solution with $\cos{(\theta_2-\theta_3)}=1, \cos{(\theta_1-\theta_2)}=\cos{(\theta_3-\theta_1)}=-1.$ For the diagonalization, we define $\psi'_i$ as
\begin{align}
&\psi'_1\equiv\frac{2}{\sqrt{6}}\psi_1+\frac{1}{\sqrt{6}}\psi_2+\frac{1}{\sqrt{6}}\psi_3,\\
&\psi'_2\equiv-\frac{1}{\sqrt{2}}\psi_2+\frac{1}{\sqrt{2}}\psi_3,\\
&\psi'_3\equiv-\frac{1}{\sqrt{3}}\psi_1+\frac{1}{\sqrt{3}}\psi_2+\frac{1}{\sqrt{3}}\psi_3.
\end{align}
Using them and (3.28), we can rewrite (3.1) as
\begin{align}
S=\int d^4x\sqrt{-g}\Big[&-\frac{1}{4}F^{\mu\nu}F_{\mu\nu}-|D_\mu\psi'_1|^2-|D_\mu\psi'_2|^2-|D_\mu\psi'_3|^2\notag\\
&-\left(m_1^2-\frac{\epsilon_{12}\epsilon_{31}}{\epsilon_{23}}\right)\psi'^2_1-\left(m_1^2-\frac{\epsilon_{12}\epsilon_{31}}{\epsilon_{23}}\right)\psi'^2_2-\left(m^2_1+\frac{\epsilon_{31}\epsilon_{23}}{\epsilon_{12}}+\frac{\epsilon_{23}\epsilon_{12}}{\epsilon_{31}}\right)\psi'^2_3\notag\\
&-\eta(\psi^4_1+\psi^4_2+\psi^4_3)\Big].
\end{align}
It is difficult to compare the free energy of this solution with others. To compare the free energy, we assume that there are only four types of the solutions with $\eta=0$:
\begin{itemize}
\item $\psi'_2\neq0, \psi'_1=\psi'_3=0$ or $\psi'_1\neq0, \psi'_2=\psi'_3=0$.
\item $\psi'_1=A\psi'_2\neq0, \psi'_3=0$ and $A$ is constant.
\item $\psi'_1=\psi'_2=0, \psi'_3\neq0$.
\item $\psi'_1=\psi'_2=\psi'_3=0$.
\end{itemize}
This assumption is natural because the mass squared of $\psi_1'$ and $\psi_2'$ is same, and the scalar fields with different mass squared do not have nonzero values together in two-scalar model \cite{Basu:2010fa} as we explained in section 2. Under this assumption, the solution with $\eta=0$ is
\begin{align}
&\psi'_1=\psi, \psi'_2=A\psi, \psi'_3=0,\\
&\psi_1=\frac{2}{\sqrt{6}}\psi, \psi_2=\frac{1-\sqrt{3}A}{\sqrt{6}}\psi, \psi_3=\frac{1+\sqrt{3}A}{\sqrt{6}}\psi,\\
&\psi>0, -\frac{1}{\sqrt{3}}<A<\frac{1}{\sqrt{3}}.
\end{align} 
Next, we check whether (3.46) is the solution with $\eta>0$. Substituting (3.28) and (3.46) to (3.25), (3.26) and (3.27), we obtain 
\begin{align}
&\nabla_\mu\nabla^\mu\psi-A_\mu A^\mu\psi-\left(m^2_1-\frac{\epsilon_{12}\epsilon_{31}}{\epsilon_{23}}\right)\psi-\frac{4}{3}\eta\psi^3=0,\\
&\nabla_\mu\nabla^\mu\psi-A_\mu A^\mu\psi-\left(m^2_1-\frac{\epsilon_{12}\epsilon_{31}}{\epsilon_{23}}\right)\psi-\frac{(1-\sqrt{3}A)^2}{3}\eta\psi^3=0,\\
&\nabla_\mu\nabla^\mu\psi-A_\mu A^\mu\psi-\left(m^2_1-\frac{\epsilon_{12}\epsilon_{31}}{\epsilon_{23}}\right)\psi-\frac{(1+\sqrt{3}A)^2}{3}\eta\psi^3=0.
\end{align}
Since these equations are not same if $\eta>0$, we conclude that (3.46) is not the solution with $\eta>0$. The same is true for $\cos{(\theta_1-\theta_2)}=1, \cos{(\theta_2-\theta_3)}=\cos{(\theta_3-\theta_1)}=-1$ and $\cos{(\theta_3-\theta_1)}=1, \cos{(\theta_1-\theta_2)}=\cos{(\theta_2-\theta_3)}=-1.$ 

The solution with $\cos{(\theta_1-\theta_2)}=\cos{(\theta_2-\theta_3)}=\cos{(\theta_3-\theta_1)}=1$ is
\begin{align}
\psi_1=\psi_2=\psi_3.
\end{align}
In fact, by substituting (3.28) and (3.51) to (3.25), (3.26) and (3.27), we obtain
\begin{align}
&\nabla_\mu\nabla^\mu\psi_1-A_\mu A^\mu\psi_1-\left(m^2_1+\frac{\epsilon_{31}\epsilon_{23}}{\epsilon_{12}}+\frac{\epsilon_{23}\epsilon_{12}}{\epsilon_{31}}\right)\psi_1-2\eta\psi_1^3=0,\\
&\nabla_\mu\nabla^\mu\psi_2-A_\mu A^\mu\psi_2-\left(m^2_1+\frac{\epsilon_{31}\epsilon_{23}}{\epsilon_{12}}+\frac{\epsilon_{23}\epsilon_{12}}{\epsilon_{31}}\right)\psi_2-2\eta\psi_2^3=0,\\
&\nabla_\mu\nabla^\mu\psi_3-A_\mu A^\mu\psi_3-\left(m^2_1+\frac{\epsilon_{31}\epsilon_{23}}{\epsilon_{12}}+\frac{\epsilon_{23}\epsilon_{12}}{\epsilon_{31}}\right)\psi_3-2\eta\psi_3^3=0,
\end{align}
and the on-shell action is
\begin{align}
S_{\textrm{on-shell}}
=\int d^4x\sqrt{-g}\Big[&-\frac{1}{4}F^{\mu\nu}F_{\mu\nu}-3|D_\mu\psi_1|^2-3\left(m^2_1+\frac{\epsilon_{31}\epsilon_{23}}{\epsilon_{12}}+\frac{\epsilon_{23}\epsilon_{12}}{\epsilon_{31}}\right)\psi_1^2-3\eta\psi_1^4\Big]\notag\\
=\int d^4x\sqrt{-g}\Big[&-\frac{1}{4}F^{\mu\nu}F_{\mu\nu}-|D_\mu\psi'|^2-\left(m^2_1+\frac{\epsilon_{31}\epsilon_{23}}{\epsilon_{12}}+\frac{\epsilon_{23}\epsilon_{12}}{\epsilon_{31}}\right)\psi'^2-\frac{\eta}{3}\psi'^4\Big],
\end{align}
where we redefine a new scalar field $\psi'$ as
\begin{align}
\psi'^2\equiv3\psi_1^2,
\end{align}
and the coefficient of the quartic potential of $\psi'$ is $\eta/3$. 
\end{enumerate}

Finally, we compare the free energy of three solutions which we found. We expect that the free energy of the solution 3a is minimum because the coefficients of the potential of $\psi'$ are smaller than those of other solutions. In order to check this expectation, we calculate the free energy by a numerical calculation. For the numerical calculation, we set $m^2_1+\frac{\epsilon_{31}\epsilon_{23}}{\epsilon_{12}}+\frac{\epsilon_{23}\epsilon_{12}}{\epsilon_{31}}=0, m_1^2-\frac{\epsilon_{12}\epsilon_{31}}{\epsilon_{23}}=-2$ and $\eta=1/2.$ Boundary conditions of the fields are
\begin{align}
\psi'(z)=\langle\mathcal{O}_2\rangle z^2\;\;(\textrm{Sol.2 and Sol.3a}),\;\;\psi'(z)=\langle\mathcal{O}_3\rangle z^3\;\;(\textrm{Sol.3b}),\;\;A_t(z)=\mu-\rho z,
\end{align}
and we fix $\mu=1$ by using the symmetry of the metric (2.1). The equations of motion can be derived from (3.31), (3.39) and (3.55).

Figure 2 is a plot of three solutions. Sol.2, Sol.3a and Sol.3b correspond to blue, red and green curves. $T_\textrm{c}$ is a critical temperature of Sol.3a. Near the critical temperature, Sol.2 and Sol.3a have similar behavior because their mass squared are same.

\begin{figure}[t]
\centering
\includegraphics[scale=0.6]{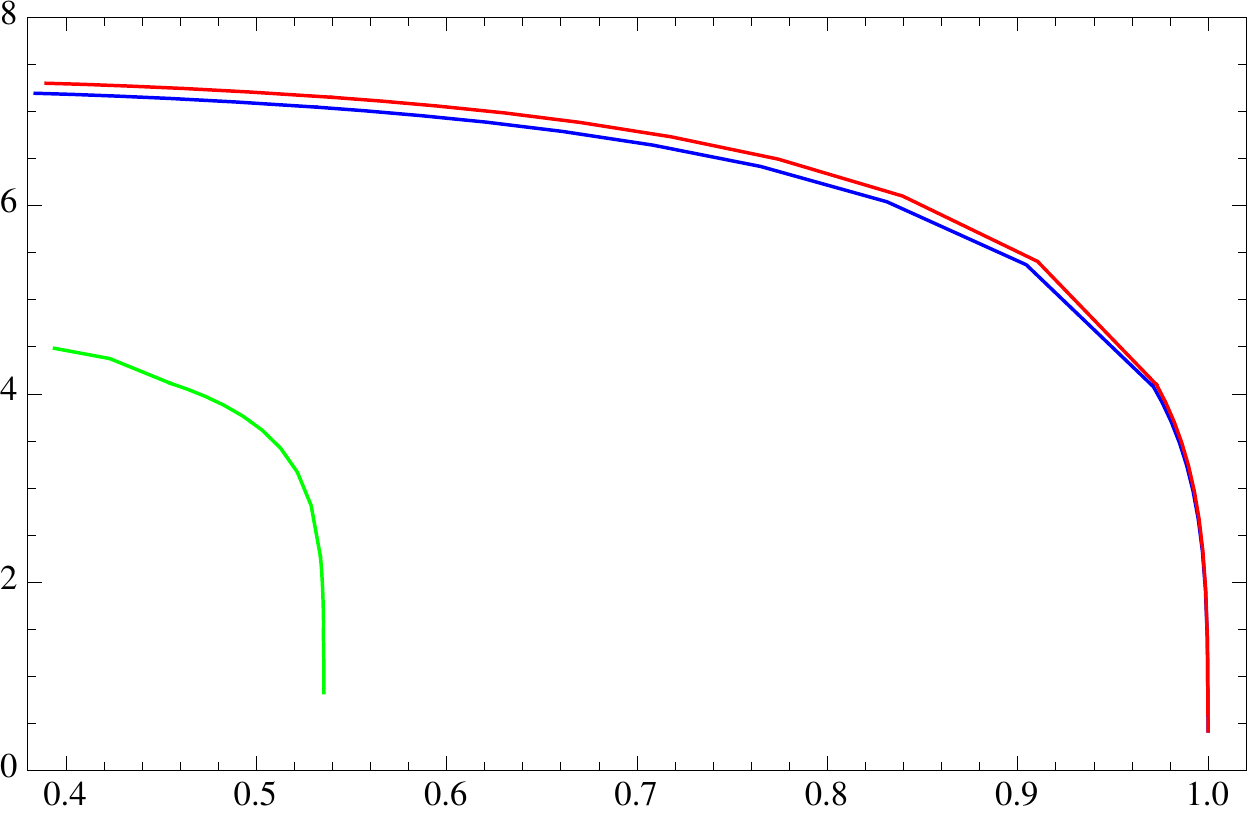}
\put(-280,120){$\langle\mathcal{O}_2\rangle^{1/2}/T_\textrm{c}$}
\put(-280,80){$\langle\mathcal{O}_3\rangle^{1/3}/T_\textrm{c}$}
\put(-0,-6){$T/T_\textrm{c}$}
\caption{Plot of the coefficients of the scalar field in each solution. Sol.2, Sol.3a and Sol.3b correspond to blue, red and green curves. $T_\textrm{c}$ is a critical temperature of Sol.3a.}
\label{fig2}
\end{figure}

Figure 3 is a plot of the free energy density. $S_{\textrm{E}}/\int dtdxdy$ corresponds to the free energy density\footnote{We ignore the Gibbons-Hawking term and a counter term because values of these terms in each solution are same in the probe limit.}, where $S_\textrm{E}$ is Euclidian action of (3.31), (3.39) and (3.55). From this figure, we conclude that the free energy of Sol.3a is minimum and the solutions (3.37) and (3.38) are chiral ground states if $\eta>0$ and (3.28) hold.

\begin{figure}[t]
\centering
\includegraphics[scale=0.6]{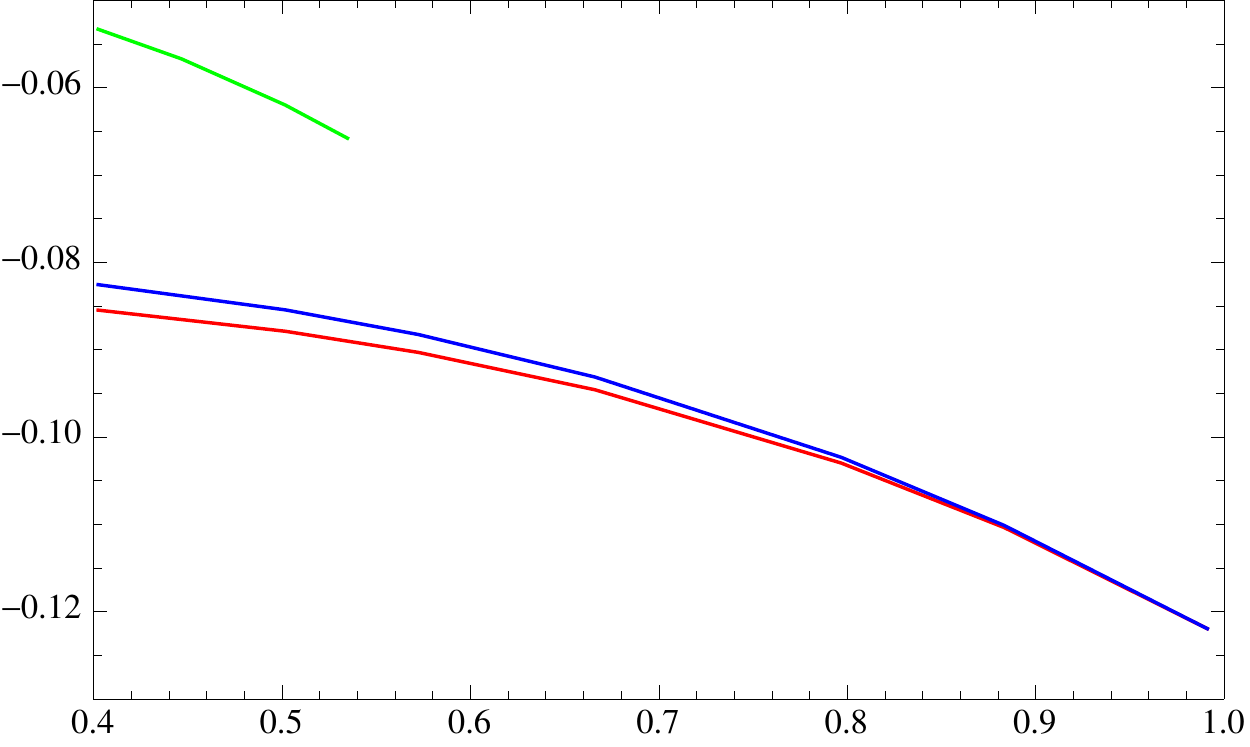}
\put(-280,120){$S_\textrm{E}/\int dtdxdy$}
\put(-0,-6){$T/T_\textrm{c}$}
\caption{Plot of the free energy density of each solution. The free energy of Sol.3a which corresponds to the red curve is minimum.}
\label{fig2}
\end{figure}

Summarizing the above, we have found the chiral ground states (3.37) and (3.38) by comparing the free energy of the 
solutions of the three-scalar model (3.1) with $\eta>0$ and (3.28). Therefore, the three-scalar model can describe the frustration in curved space-time and introduce chirality as an order parameter. 
\section{Summary and discussion}
\label{sec4}

In this paper, we have analyzed the holographic superconductor model with some scalar fields and the Josephson coupling from the view point of gauge/gravity correspondence. We have seen that the frustration does not occur in the holographic two-scalar model because there is one Josephson coupling term only. On the other hand, we have found that there are several solutions whose free energy is same by the frustration in the holographic three-scalar model with three Josephson coupling terms if $\eta=0$, $\epsilon_{12}\epsilon_{23}\epsilon_{31}>0$ and (3.18) hold. Furthermore, we have analyzed the free energy of the solutions with $\eta>0$. We have found that there are the solutions (3.37) and (3.38) which correspond to the chiral ground states with (3.28).

To compare the free energy of the solutions, we reduced the solutions of the three-scalar model to those of the one-scalar model. It is important to check whether there are other solutions which do not reduce to those of the one-scalar model. Moreover, analysis with a back reaction is also important. 

For anomalous Hall effect from chirality, a coupling between an electron and localized spins is important \cite{Ohgushi}. Therefore, the three-scalar model with fermions coupled with complex scalar fields may have strange property.

In Ref.~\cite{Tanaka}, the existence of a chiral domain wall is discussed in the context of condensed matter physics. It is interesting to check whether there is a solution which corresponds to the chiral domain wall in a holographic model. In this paper, we considered specific values of the parameters as (3.28). Comparing the free energy with other values of the parameters which satisfy (3.18) is also interesting because there is some possibility of a phase transition by changing the parameters. They are future directions of this study. 

\acknowledgments

I would like to thank K. Hashimoto for helpful discussion, careful reading of the manuscript and useful suggestion about section 3.2. I would also like to thank H. Kawamura and A. Sonoda for useful comment.

%%%%%%%%%%%%%%%%%%%%%%%%%%%%%%%%%%%%%%%%%

\end{document}